# SEE++: Evolving Snowpark Execution Environment for Modern Workloads

Gaurav Jain, Brandon Baker, Joe Yin, Chenwei Xie, Zihao Ye, Sidh Kulkarni,
Sara Abdelrahman, Nova Qi, Urjeet Shrestha, Mike Halcrow, Dave Bailey, Yuxiong He

*Snowflake, Inc*

*Abstract*— Snowpark enables Data Engineering and AI/ML workloads to run directly within Snowflake by deploying a secure sandbox on virtual warehouse nodes. This Snowpark Execution Environment (SEE) allows users to execute arbitrary workloads in Python and other languages in a secure and performant manner. As adoption has grown, the diversity of workloads has introduced increasingly sophisticated needs for sandboxing. To address these evolving requirements, Snowpark transitioned its in-house sandboxing solution to gVisor, augmented with targeted optimizations. This paper describes both the functional and performance objectives that guided the upgrade, outlines the new sandbox architecture, and details the challenges encountered during the journey, along with the solutions developed to resolve them. Finally, we present case studies that highlight new features enabled by the upgraded architecture, demonstrating SEE's extensibility and flexibility in supporting the next generation of Snowpark workloads.

*Keywords—Security Sandboxing, Data Engineering, AI/ML, Serverless Compute, High Performance Compute*

## I. Introduction

Snowflake AI Data Cloud [1] introduced Snowpark [2] as a managed turnkey solution that supports Data Engineering and AI/ML workloads using Python and other programming languages. Snowpark provides the DataFrame APIs to accept operators defined in non-SQL languages and emit SQL queries to execute in Snowflake virtual warehouse. Additionally, Snowpark also built a secure sandbox based on syscall filtering to allow user code to be pushed down to execute performantly and securely in Snowpark Execution Environment.

Since the introduction of Snowpark, we see strong adoption momentum. With the rapid growth of adoption, and the wide variety of newly emerged use cases, relying on legacy sandbox's syscall filtering configurations to block malicious actions is not sustainable. And in extreme cases, customer workloads may require certain specific syscalls which are dangerous to allow to the kernel. We therefore realized that the existing Snowpark sandbox solution is no longer feasible to meet the modern workloads' needs, and a redesign was in order. We focused on three objectives as we redesign Snowpark sandbox solution:

- Functionality Compatibility - With more workloads coming to Snowpark, the new sandbox solution needs to be compatible for them. Specifically, the sandbox should be able to execute arbitrary Python packages referenced in those workloads, given the popularity of Python in Data Engineering and AI/ML space.
- Performance - Snowpark attracts new workloads and retains existing workloads by its advantage over price/performance. We should keep this advantage when introducing the new sandbox solution.
- Maintainability - Since customers' workloads are evolving and hard to predict, we expect the sandbox solution can naturally adapt to emerging requirements without manual intervention.

To achieve these objectives, we adopted gVisor [3, 4] as the foundation for the new Snowpark sandbox solution. gVisor is an industry-wide, widely adopted sandbox, which focuses on security, efficiency and ease of use. Secondly, we introduced a standardized base image encapsulating common system-level dependencies required by a broad range of Python packages, for the purpose of dependency management. Given Snowpark's design to support arbitrary workload execution and its novel architecture that colocates the sandbox with Snowflake's query engine for SQL workloads processing, we faced distinctive challenges. To address these, we developed targeted gVisor enhancements focused on sandbox compatibility and memory management.

The rest of this paper is organized as follows. Section 2 covers the background of legacy sandbox solution. Section 3 presents the architecture of upgraded modern sandbox. We then discuss the challenges we encountered along with solutions we built in Section 4, and further product features unblocked by the upgrade in Section 5. Finally, Section 6 concludes the paper.

## II. Background – Legacy Sandbox Overview

Snowpark [2] provides the execution environment to run arbitrary user code in Snowflake virtual warehouses. Since non-SQL user code is more flexible and could be malicious, a security boundary is critically essential to isolate the code.

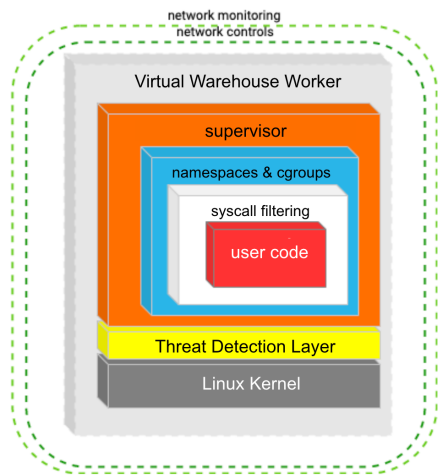

Fig. 1. Legacy Snowpark Sandbox

The legacy sandbox relies on syscall filtering to provide security enforcement. Although this is sufficient from security perspective, the syscall filtering configurations require regular revisiting based on logging and insights captured by the supervisor process to make adjustments. This is inefficient, error-prone, and in extreme cases where customer workloads need to access a few dangerous syscalls, it's challenging to support relevant use cases.

## III. MODERN SNOWPARK SANDBOX ARCHITECTURE

The modern Snowpark sandbox builds on top of the widely adopted gVisor [3, 4] implementation to allow flexible syscall invocations, and defines a base image to contain the popular and essential system dependencies to power vast customer workloads in Snowpark.

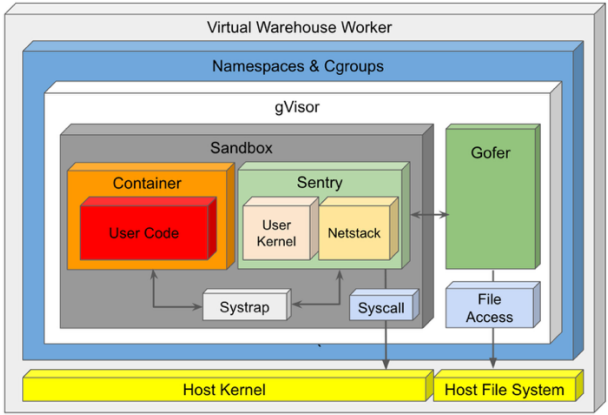

Fig. 2. Modern Snowpark Sandbox

### A. gVisor for Sandboxing

gVisor is a secure sandbox solution developed by Google that provides strong isolation for workloads by intercepting and emulating the majority of essential Linux system calls. Unlike traditional sandboxing runtime solutions that rely on the host kernel, gVisor executes most kernel functionalities in user space, significantly reducing the host's attack surface. gVisor consists of below key components:

- Sentry: The user-space kernel that implements much of the Linux syscall interface in Go. Sentry emulates the kernel behaviors and handles syscalls intercepted from sandboxed workloads.
- Gofer: A process that mediates filesystem access between the sandbox and the host using the 9P protocol [5]. It allows gVisor Sentry to perform file operations without direct host kernel involvement.
- Systrap: A modern platform backend that replaces the previous ptrace mechanism. It uses seccomp-bpf and userfaultfd to trap syscalls and manage memory with significantly lower overhead compared to the ptrace implementation [6].

Snowpark adopts gVisor as the modern sandbox solution for running non-SQL code from stored procedures and user defined functions to contain potential malicious intentions.

### B. Sandbox Base Image

In addition to sandboxing, Snowpark isolates user code by introducing a chroot directory, which is a minimal file system that serves as the root environment for the sandbox. This directory contains the essential binaries and libraries required to support the execution of user code.

To support modern workloads hosting needs in Snowpark, we standardized chroot directory setup by replacing it with a pre-defined base image, which captures essential binaries and libraries to power Snowpark runtime and customers' user code. At sandbox startup, since gVisor is an OCI-compatible [7] container runtime, the sandbox instance is bootstrapped from the base image to include system-level dependencies.

Sandbox base image provides the option for standardizing the runtime environment, and allows Snowpark to decouple user code dependencies from host dependencies installed on Snowflake virtual warehouses for further extensibility and flexibility.

### C. Modern Snowpark Sandbox Advantages

The modern Snowpark sandbox architecture adopts gVisor as the foundation for sandboxing solution and builds upon pre-defined base image for bootstrapping the standard and extensible runtime environment. This architecture brings multiple advantages to address legacy sandbox solution pain points.

**Functionality Compatibility**: Unlike the legacy sandbox solution where it maintains syscall filtering configurations, gVisor implements the majority of essential syscalls in user space, allowing modern workloads to run seamlessly leveraging those implementations. In addition, the sandbox base image defines the standardized runtime dependencies to further boost the functionality compatibility.

**Performance**: gVisor brings the performance advantage for sandboxed environments by minimizing host kernel interaction through syscall emulation in user space. Its modern backend, systrap, uses mechanisms like seccomp-bpf and userfaultfd to intercept syscalls efficiently, reducing overhead and enabling better scalability for workloads. This design guarantees low-latency user code execution, while maintaining strong isolation for security enforcement.

**Maintainability**: While the legacy sandbox solution needs regular reviews on syscall filtering logging to understand the user workloads' evolvements and make syscall allowlisting adjustment, gVisor's nature of implementing syscall in user space avoids such syscall filtering configuration maintenance overhead.

Along the journey of upgrading the Snowpark Execution Environment, not everything works out of the box. We encountered many challenges, both from functionality and performance perspectives, and developed solutions for addressing them. Some representing ones are showcased in the next section.

## IV. CHALLENGES AND RESOLUTIONS

As Snowpark transitioned to a modern sandbox architecture, migrating existing workloads introduced significant challenges. These included frequent workload crashes and noticeable performance degradations. To diagnose and mitigate those problems, we employed a variety of low-level tools. These included *strace* for analyzing user space processes, *perf* for Linux kernel profiling, and utilities such as *dd* and *sar* for disk I/O profiling. These mechanisms provided visibility into bottlenecks and failure modes within the system. Based on these insights, we implemented several key optimizations. Examples include tuning Linux Transparent Huge Pages (THP), adjusting gVisor's copy-on-write memory sizing on page faults, and refining heuristics for file descriptor table resizing. While many of these optimizations remain internal to the Snowpark stack, others were developed within gVisor itself and contributed back to the open-source community. In the following sections, we highlight two of these contributions and the broader lessons they reveal.

*A. Virtual Memory Management Optimization*

As part of the effort to integrate gVisor as the secure sandbox into Snowpark Execution Environment, we observed a severe mismatch in virtual memory area (VMA) [8] allocation behavior between gVisor and the native Linux kernel. A representative memory-intensive Python workload that created a few hundreds of VMA entries under Linux resulted in over 500 times more VMA entries when executed in gVisor. This type of discrepancy caused the workload to exceed the default 65,530 VMA limit in Linux kernels and crashed the sandbox.

The root cause was identified as a mismatch in memory allocation direction between gVisor's virtual address space and the file offset allocation in its memfd-backed mappings. Specifically, when a VMA lacked a last faulted address, which is used by gVisor to infer access direction, the allocator defaulted to bottom-up file offset allocation, even though the address space was allocated top-down. This misalignment prevented the Linux kernel from coalescing adjacent VMAs, leading to fragmentation and excessive VMA count. Additionally, gVisor's in-memory VMA merge logic could drop last faulted address during merges, compounding the problem by further preventing correct allocation direction inference. The proposed approach aligns the file offset allocation direction with the actual address space growth direction and preserves last faulted address during merges.

This optimization significantly reduces the number of VMA entries generated during memory-heavy workloads, improving kernel compatibility, avoiding configuration workarounds, and supporting scalability to larger memory footprints, such as Snowpark optimized Warehouses, which provide higher memory availability to users. It also reduces potential kernel overhead in managing large VMA trees, contributing to more efficient and predictable performance under gVisor.

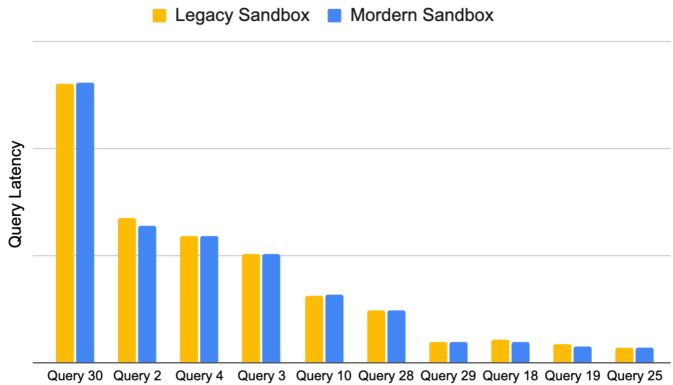

Fig. 3. TPCx-BB Query Latency Comparison

We validated this memfd offset allocation direction adjustment approach by synthetic benchmark generated based on the observed relevant production workloads. Specifically, it captures the common pattern of repeatedly appending new lists into an existing list to build a two-dimensional array. This is a popular pattern where users are preparing a two-dimensional array as a DataFrame to fit into packages, such as pandas [9], scikit-learn [10], etc, for further data processing. Our approach effectively reduced VMA entries in gVisor runtime by 182 times in this benchmark query, while achieving minor performance difference compared to the legacy sandbox runtime. To further evaluate modern sandbox's performance impact, we leveraged TPCx-BB [11] to benchmark both sandbox solutions. Figure 3 visualizes the top ten longest queries' latency comparison, and overall we observed 1.5% performance improvement when running full benchmarks on the modern sandbox architecture compared to the legacy architecture.

*B. ELF Loader Semantics Optimization*

As part of validating gVisor for secure sandboxing of data processing workloads, we identified a critical compatibility issue affecting the execution of certain Executable and Linkable Format (ELF) binaries under gVisor. Specifically, the issue surfaced when running Python workloads using the prophet [12] package. The workload triggered a segmentation fault due to a mismatch in how gVisor handled ELF memory layout compared to the native Linux kernel. The root cause was traced to gVisor's mishandling of certain special ELF binaries in which the DYNAMIC section resided outside of all declared LOAD segments, but still within the extended memory range created by page alignment, illustrated in Figure 4.

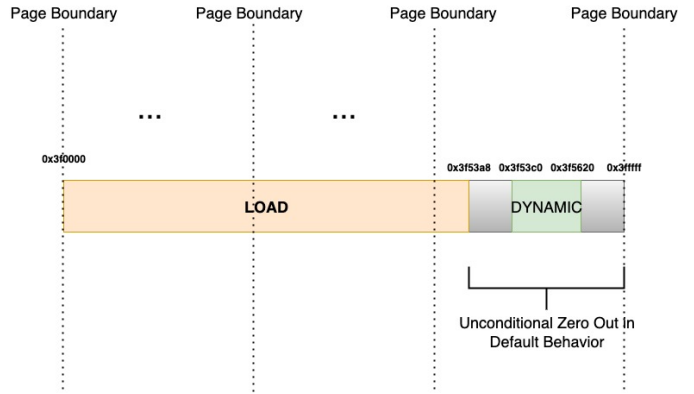

Fig. 4. DYNAMIC section resides outside of LOAD segments

In ELF program headers, FileSiz denotes the number of bytes a segment occupies in the file, while MemSiz specifies the number of bytes it will occupy in memory once loaded. In native Linux, when MemSiz is greater than FileSiz, the kernel zeroes out only the additional memory beyond FileSiz as prescribed in the ELF program header. However, gVisor unconditionally zeroed out the full page-aligned extension of the LOAD section regardless of whether it was explicitly covered by a LOAD directive. This resulted in corruption of metadata in certain Python libraries, where critical sections such as DYNAMIC were implicitly expected to be present within that memory region.

To resolve this observed incompatibility, we introduced an optimization in gVisor that modifies its memory loader behavior to match Linux semantics, which zeros out only the portion of memory explicitly indicated by the ELF header, when MemSiz is greater than FileSiz. This alignment with the Linux kernel's behavior enables successful execution of previously failing workloads and improves compatibility with a broader range of dynamically linked ELF executables.

This optimization enhances gVisor's utility as a secure Linux syscall emulator in enterprise environments by reducing silent incompatibilities in real-world Python workloads, while maintaining the sandbox's overall integrity and correctness.

## V. Modern Sandbox Unblocked Snowpark Features

The upgraded Snowpark sandbox enhances functionality compatibility, performance, and maintainability, creating a foundation for broader feature enablement to enrich the Snowflake product. This section presents two key product features that rely on the modern sandbox architecture.

### A. Snowpark in Snowflake Serverless Tasks

Snowflake Serverless Tasks [13] provide a fully managed, event-driven execution model for automating user workloads. By abstracting away the underlying infrastructure, Snowflake dynamically allocates compute resources based on workload characteristics and executes workloads in a multi-tenant setup. This model improves operational efficiency as well as ease of use for customers.

The integration of gVisor as the sandboxing solution is a key enabler for extending Serverless Tasks support to power Snowpark stored procedures and user defined functions. gVisor's user space syscall interception provides stronger process isolation than the legacy sandbox solution. This security upgrade guarantees the safe execution of untrusted user code in a multi-tenant environment.

### B. Snowflake Artifact Repository

Snowpark allows customers to run a wide variety of Data Engineering and AI/ML workloads in Python by providing the accessibility to various Python packages. Snowflake Artifact Repository [14] is the critical component that is introduced to allow users to reference any PyPI packages in Snowpark Python stored procedures and user defined functions.

The modern Snowpark sandbox architecture plays a key role to unblock Snowflake Artifact Repository. gVisor's user space syscall implementation makes sure packages' arbitrary syscalls are functioning, and we don't need to maintain or update any syscall handling mechanism or configurations to support newly emerging packages. Furthermore, the sandbox base image provides a well-defined runtime environment to host common system libraries for packages' dependency needs.

## VI. Conclusion

In summary, Snowpark Execution Environment evolves by adopting a modern sandbox architecture that uses gVisor as the underlying sandboxing solution and defines the base image to contain essential system dependencies as the stable runtime environment. This upgrade brings functional compatibility to allow arbitrary workloads' execution, high performance to reinforce price/performance advantages of Snowpark, and ease of maintenance that the sandbox environment can naturally adapt to rapidly changing workload characteristics. These enhancements elevate Snowpark's capabilities to power diverse Data Engineering and AI/ML workloads, and open doors for further extensions to embrace more advanced features in the future.

### Acknowledgment

Upgrading the Snowpark Execution Environment is the result of the hard work and dedication of numerous talented individuals. We extend our gratitude to the entire Snowpark and partner teams for their remarkable support along the journey. It is an honor and a privilege to work with such an outstanding team, and we are constantly inspired by their excellence.